\documentclass[12pt, prb]{revtex4}

\usepackage{amsmath}
\usepackage{graphicx}

\begin{document}

\begin{titlepage}
\title{Effect of charge distribution on the translocation of an inhomogeneously charged polymer through a nanopore}
\author{Aruna Mohan}
\author{Anatoly B. Kolomeisky}
\author{Matteo Pasquali}
\affiliation{Department of Chemistry and Department of Chemical and Biomolecular Engineering, Rice University,
Houston, TX 77005}
\date{\today}

\begin{abstract}
We investigate the voltage-driven translocation of an inhomogeneously charged polymer through a nanopore by utilizing discrete and continuous stochastic models. As a simplified illustration of the effect of charge distribution on translocation, we consider the translocation of a polymer with a single charged site in the presence and absence of interactions between the charge and the pore. We find that the position of the charge that minimizes the translocation time in the absence of pore--polymer interactions is determined by the entropic cost of translocation, with the optimum charge position being at the midpoint of the chain for a rodlike polymer and close to the leading chain end for an ideal chain. The presence of attractive or repulsive pore--charge interactions yields a shift in the optimum charge position towards the trailing end and the leading end of the chain, respectively. Moreover, our results show that strong attractive or repulsive interactions between the charge and the pore lengthen the translocation time relative to translocation through an inert pore. We generalize our results to accommodate the presence of multiple charged sites on the polymer. Our results provide insight into the effect of charge inhomogeneity on protein translocation through biological membranes.
\end{abstract}

\maketitle
\end{titlepage}

\section{\label{sec:intro} Introduction}

The migration of biopolymers such as DNA, RNA and proteins through nanopores plays an important role in several biological processes \cite{lodish}. In particular, the transport of proteins from the ribosome, where they are synthesized, to specific locations within or outside the cell occurs via protein translocation across a variety of biological membranes. The targeting of a protein to a membrane is typically mediated by a protein--RNA complex known as the signal recognition particle when translocation occurs concurrently with protein synthesis at the ribosome, or by chaperones or protein complexes when translocation occurs subsequent to protein synthesis. In both cases, the presence of a sequence of amino acids near either the amino-terminus or the carboxyl-terminus of the protein, known as the signal sequence, allows the recognition of the protein by the targeting machinery, which subsequently delivers the protein to the translocation channel. The protein must then be moved unidirectionally across the channel \cite{eichler}. The driving force for protein translocation comes from ATP hydrolysis or the presence of an electrochemical potential difference across the membrane \cite{schatz1, schatz2, schatz3}.

Several experimental studies have revealed that protein translocation across biological membranes is dependent on the charge distribution of the residues comprising the protein \cite{andrews, puziss, geller, kajava, efremov, vergunst}. The influence of charge distribution on translocation has been variously attributed to possible changes in protein conformation or orientation under an applied potential difference \cite{geller, efremov} or specific interactions with the export machinery \cite{vergunst}. Moreover, charged peptides have technological applications in drug delivery into cellular compartments \cite{deshayes}. It has been suggested that the presence of positive charges in cell-penetrating peptides such as penetratin permits electrostatic interactions with membrane phospholipid groups, thereby facilitating the cellular internalization of the peptide \cite{deshayes}. However, the role played by charge inhomogeneity in determining the rate of translocation remains unclear.

Whereas the translocation of uniformly charged nucleic acids or proteins has been the subject of many recent experimental, computational and theoretical investigations \cite{kasianowicz, sung, degennes, muthu, lubensky, henrickson, mellerprl, kardar1, ambjornsson, meller, slonkina, metzler, kardar2, makarov1, makarov2, storm, dekker, matysiak, kong}, the effect of charge inhomogeneity on protein translocation is relatively unexplored. In view of the biological and technological relevance of protein translocation, we here investigate the effect of charge inhomogeneity on the voltage-driven translocation of nonuniformly charged polymers. The paper is organized as follows. In Sec. \ref{sec:onecharge}, we treat the translocation of a simplified model polymer bearing a single charge through a nanopore under an applied voltage difference by means of discrete and continuous random walk models. We consider ideal, self-avoiding and rodlike chains, and demonstrate that the charge position leading to the minimum time of translocation through an inert pore is governed solely by the nature of the polymer or, equivalently, the solvent quality. The effect of attractive or repulsive interactions between the charged site and the nanopore is also considered. Section \ref{sec:twocharge} extends our analysis to a polymer carrying multiple charges, and presents explicit results for the translocation of a polymer possessing two charged sites through an inert pore. Finally, Sec. \ref{sec:sum} provides a summary of our findings. 

\section{\label{sec:onecharge} Model}

\begin{figure}
\centering
\resizebox{83mm}{!}{\includegraphics{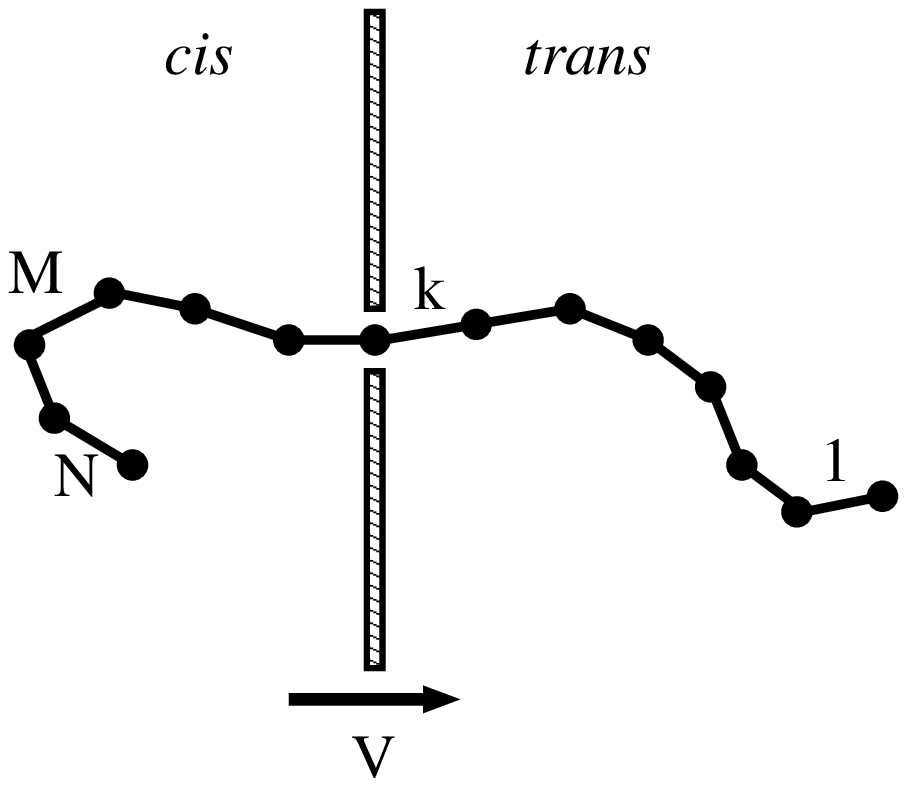} } 
\caption{A translocating chain composed of $N$ segments carrying a single charge of magnitude $q$ at the location of the $M^\mathrm{th}$ segment from the leading end. The number of translocated segments is denoted by $k$. The electric potential vanishes on the \emph{cis} side and has a non-zero value, denoted by $V$, on the \emph{trans} side. The direction of translocation, which coincides with the direction of increasing electric potential, is indicated by the arrow. \label{fig:transloc}}
\end{figure}

We consider the translocation of a polymer comprised of $N$ segments through a nanopore in an infinite, planar membrane, as illustrated in Fig. \ref{fig:transloc}. The nanopore is assumed small enough that only a single monomer may be accommodated at a time, and hairpin configurations are disallowed from occurring. The electric potential on the left (\emph{cis}) side of the membrane is assumed to vanish, while the potential takes the value $V \neq 0$ on the right (\emph{trans}) side of the membrane. The polymer chain is assumed to possess a single charge of magnitude $q$ at the location of the $M^\mathrm{th}$ segment from the leading end. The pore may interact with the charged segment when the latter is situated immediately adjacent to the pore on the \emph{cis} side. Pore--charge interactions are quantified by the interaction energy parameter $\epsilon$, which is positive when the net interaction is repulsive, and negative in the case of net pore--charge attraction. At the start of the translocation process, the leading monomer is assumed to be located adjacent to the pore on the \emph{cis} side. Thus, only successful translocation events are considered.

We model the translocation process as being equivalent to the diffusion of the translocation coordinate, namely, the number of polymer segments that have been transported across the membrane, over a free energy barrier \cite{muthu}. Although the approach adopted by us cannot capture anomalous dynamics \cite{kardar1, kardar2, panja} stemming from memory effects in chain tension across the pore \cite{panja}, we expect our model to adequately capture the physical mechanisms by which charge inhomogeneity influences translocation dynamics.

The instantaneous free energy of the polymer chain when $k$ segments have been translocated may be written as the sum of the free energies of the portions of the chain on the \emph{cis} side and \emph{trans} side containing $N-k$ and $k$ segments, respectively, whereby we obtain the expression
\begin{equation}
\beta F_k = 
\begin{cases}
(1-\gamma) \ln \left[ k\left(N-k\right)\right] \text{ for } 1 \leq k \leq M-2, \\
(1-\gamma) \ln \left[ k\left(N-k\right)\right] + \beta\epsilon \text{ for } k=M-1, \\
(1-\gamma) \ln \left[ k\left(N-k\right)\right] - \beta qV \text{ for } M \leq k \leq N-1, \label{eq:betaF}
\end{cases}
\end{equation}
where $\beta=1/(k_B T)$ and $\gamma$ is the exponent determining the number of configurations $Z_N \sim N^{-\left( 1 - \gamma \right) }$ of a wall-tethered polymer of $N$ segments \cite{dimarzio, eisen, muthu}. The parameter $\gamma$ takes the value $0.5$, $0.69$ and $1$ for an ideal random walk, a self-avoiding walk and a rodlike chain, respectively \cite{dimarzio, eisen}. Furthermore, at the start of the translocation process ($k=0$), the free energy takes the form
\begin{equation}
\beta F_0 = 
\begin{cases}
(1-\gamma) \ln N \text{ if } M \neq 1, \\
(1-\gamma) \ln N + \beta\epsilon \text{ if } M=1, \label{eq:betaFk0}
\end{cases}
\end{equation}
whereas at the end of the translocation process ($k=N$),
\begin{equation}
\beta F_N = (1-\gamma) \ln N - \beta qV. \label{eq:betaFkN}
\end{equation}

The transport of the polymer through the pore may be modeled as the discrete random walk of the translocation coordinate $k$, whereby the probability $P_k(t)$ of having a configuration of $k$ translocated segments at time $t$ is governed by the Master equation \cite{vankampen}
\begin{equation}
\frac{\partial P_k(t)}{\partial t} = u_{k-1} P_{k-1} + w_{k+1} P_{k+1} -\left( u_k + w_k \right) P_k. \label{eq:master}
\end{equation}
In Eq. (\ref{eq:master}), the terms $u_k$ and $w_k$ refer, respectively, to the forward transition rate from a configuration of $k$ to one of $k+1$ translocated segments and the reverse transition rate from a configuration of $k$ to one of $k-1$ translocated segments. Equation (\ref{eq:master}) must be solved subject to a reflecting boundary condition at $k=0$ and, concomitantly, an absorbing boundary condition at $k=N$. Although more general boundary conditions may be considered, the absorbing boundary at $k=N$ is chosen for consistency with experimental measurements, wherein only successful translocation events are recorded. The forward and reverse transition rates are related by the detailed balance condition:
\begin{equation}
\frac{w_{k+1}}{u_{k}} = \exp[\beta\left(F_{k+1} - F_{k} \right) ]. \label{eq:detailedbal}
\end{equation}
We may further introduce the parameter $\theta$ specifying the distribution of the free energy difference expressed in Eq. (\ref{eq:detailedbal}) between the forward and reverse transition rates, yielding the expressions \cite{kotsev}
\begin{equation}
u_k = D \exp[-\theta \beta\left(F_{k+1} - F_{k} \right) ] \label{eq:uk}
\end{equation}
and 
\begin{equation}
w_{k+1} = D \exp[\left( 1-\theta \right) \beta\left(F_{k+1} - F_{k} \right) ], \label{eq:wk}
\end{equation}
where $D$ is a constant having the units of inverse time. The parameter $\theta$ quantifies the distance between the transition state and the reactant, namely, the state of $k$ translocated segments, on a reaction coordinate diagram. Consequently, $\theta=0.5$ represents the situation wherein the transition state is symmetric with respect to the reactant (having $k$ translocated segments) and product (having $k+1$ translocated segments). Additionally, $\theta$ specifies the distribution of the free energy difference between the forward and reverse transition rates.

The translocation time may now be obtained as the time of first passage to the absorbing boundary $k=N$. For the discrete random walk described by Eq. (\ref{eq:master}), the first passage time $\tau$ is given by the expression \cite{vankampen, pury}
\begin{equation}
\tau =\sum_{k=0}^{N-1}\frac{1}{u_{k}} + \sum_{k=0}^{N-2}\frac{1}{u_{k}}
\sum_{i=k+1}^{N-1} \prod \limits_{j=k+1}^{i} \frac{w_{j}}{u_{j}}. \label{eq:fptdiscrete}
\end{equation}
The above expression is readily evaluated upon combining Eqs. (\ref{eq:uk}) and (\ref{eq:wk}), specifying the transition rates, with the free energy expressions given by Eqs. (\ref{eq:betaF})--(\ref{eq:betaFkN}).

\begin{figure}
\centering
\resizebox{83mm}{!}{\includegraphics{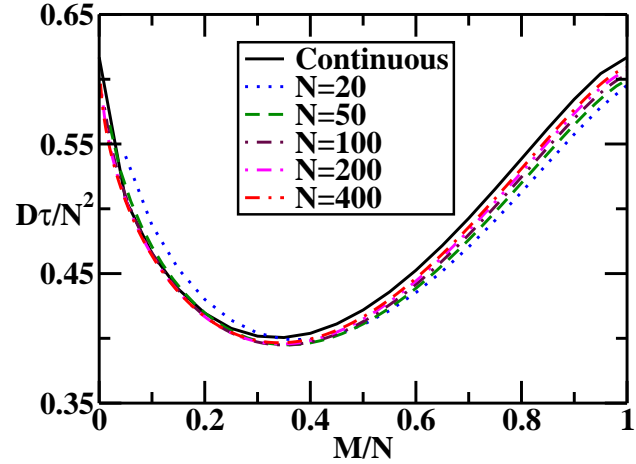} } 
\caption{Rescaled translocation time $D\tau/N^2$ as a function of the charge position expressed as a fraction of the chain length, $M/N$, for ideal chains of several lengths with $\beta qV=1$, $\beta\epsilon=0$ and $\theta=0$. Also shown for comparison is the translocation time obtained in the continuous limit of the chain. \label{fig:Ncollapse} }
\end{figure}

\begin{figure}
\centering
\resizebox{83mm}{!}{\includegraphics{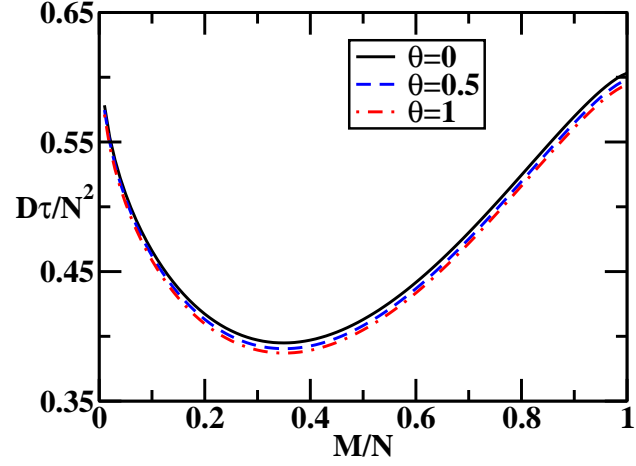} } 
\caption{Rescaled translocation time $D\tau/N^2$ as a function of the fractional charge position $M/N$ for an ideal chain of $N=100$ segments at $\beta qV=1$ and $\beta\epsilon=0$, with $\theta=0$ (solid line), $\theta=0.5$ (dashed line) and $\theta=1$ (dashed-dotted line). \label{fig:theta}}
\end{figure}

We first consider translocation through an inert pore in the absence of pore--charge interactions, i.e., $\epsilon=0$. Figures \ref{fig:Ncollapse} and \ref{fig:theta} illustrate our results for ideal chains. Similar plots (not shown here) are obtained for self-avoiding and rodlike chains.  Figure \ref{fig:Ncollapse} demonstrates the collapse of data for different chain lengths when the scaled translocation time $D\tau/N^2$ is plotted as a function of the position of the charge from the leading chain end expressed as a fraction of the chain length, $M/N$. The scaling of $\tau$ with $N^2$ is consistent with the scaling observed for translocation over an entropic barrier \cite{muthu}. Surprisingly, the minimum translocation time occurs at $M/N \simeq 0.3$ for all values of $N$ from Fig. \ref{fig:Ncollapse} with $\beta qV=1$ and $\theta=0$, although the naive expectation may be that the minimum occurs at $M/N=0.5$.  The corresponding results for self-avoiding and rodlike chains are similar, but the minimum in the translocation time occurs at $M/N \simeq 0.4$ for a self-avoiding chain and at $M/N=0.5$ for a rodlike chain. It should be noted, however, that upon averaging the translocation times obtained when the charge is located at monomer positions $M$ and $N-M$ from the leading end, the resulting symmetric mean translocation time exhibits a minimum at the midpoint $M/N=0.5$. Nonetheless, owing to the fact that one end of a protein is preferentially delivered to the pore during cellular transport, the two ends are distinguishable and charges at positions $M$ and $N-M$ do not behave identically.

The determination of the position of the charge at which the translocation time is minimized is facilitated upon taking the continuous limit of the discrete Master equation, (\ref{eq:master}), under the assumption that the forward transition rates $u_k$ are constant, i.e., $\theta=0$. While the parameter $\theta$ may, in general, assume any value from $0$ to $1$, it is evident from Fig. \ref{fig:theta} that varying $\theta$ has little effect on the translocation time. Consequently, we expect the results to be largely unaffected upon setting $\theta=0$ and taking the continuous limit of Eq. (\ref{eq:master}). In the subsequent development, we continue to set $\theta=0.5$ in the discrete model corresponding to a symmetric reaction coordinate diagram \cite{kotsev}, although the continuous limit is based on the assumption $\theta=0$. The continuous limit of Eq. (\ref{eq:master}) yields the equation \cite{vankampen, muthu}
\begin{equation}
\frac{\partial p}{\partial t} = \beta D \frac{\partial}{\partial k} \left(\frac{\partial F}{\partial k}p \right) + D\frac{\partial^2 p}{\partial k^2},
\end{equation}
where $p(k,t)$ is the probability density of the translocation coordinate $k$, which is now allowed to vary continuously between $0$ and $N$. The corresponding mean first passage time to the absorbing boundary at $k=N$ is \cite{vankampen, muthu}
\begin{equation}
\tau =\int_{0}^{N-1}dk^{\prime }e^{\Phi (k^{\prime })}\int_{0}^{k^{\prime
}}dk^{\prime \prime }\frac{e^{-\Phi (k^{\prime \prime })}}{D}, \label{eq:taucont}
\end{equation}
with
\begin{equation}
\Phi \left( k\right) = \int_{0}^{k}dk^{\prime }\beta \frac{\partial
F_{k^{\prime }}}{\partial k^{\prime }}=\beta \left(F_{k} - F_{0} \right). \label{eq:Phi}
\end{equation}
Equations (\ref{eq:taucont}) and (\ref{eq:Phi}) may be combined with Eqs. (\ref{eq:betaF})--(\ref{eq:betaFkN}), upon setting $\epsilon=0$ and replacing the upper limit of the outer integral on the right hand side of Eq. (\ref{eq:taucont}) with $N$ in the limit of large $N$, yielding the following expression for the translocation time in the continuous limit:
\begin{multline}
D\tau =\int_0^N dk^\prime \left[ k^{\prime }\left( N-k^{\prime }\right) %
\right] ^{1-\gamma }\int_{0}^{k^{\prime }}dk^{\prime \prime }\left[
k^{\prime \prime }\left( N-k^{\prime \prime }\right) \right] ^{-\left(
1-\gamma \right) }- \\
\left(1-e^{-\beta qV} \right)\int_{M}^{N}dk^{\prime }\left[ k^{\prime }\left( N-k^{\prime }\right) \right]
^{1-\gamma } \int_{0}^{M}dk^{\prime \prime } \left[
k^{\prime \prime }\left( N-k^{\prime \prime }\right) \right] ^{-\left(
1-\gamma \right)}. \label{eq:Dtau}
\end{multline}
Upon rearranging, Eq. (\ref{eq:Dtau}) yields the expression
\begin{multline}
\frac{D\tau }{N^{2}}=\int_{0}^{1}dy\frac{B(y,\gamma ,\gamma )}{\partial
B(y,\gamma ,\gamma )/\partial y}- \\
\left( 1-e^{-\beta qV}\right) \left[
B\left( 2-\gamma ,2-\gamma \right) -B\left( \frac{M}{N},2-\gamma ,2-\gamma
\right) \right] B\left( \frac{M}{N},\gamma ,\gamma \right), \label{eq:DtaubNsq}
\end{multline}%
where we have introduced the incomplete beta function
\begin{equation}
B\left( x,a,b\right) =\int_{0}^{x}dx^{\prime }x^{\prime a-1}\left(
1-x^{\prime }\right) ^{b-1}. 
\end{equation}%

Equation (\ref{eq:DtaubNsq}) corroborates the scaling $\tau \sim N^2/D$ observed to occur from the discrete solution. Figure \ref{fig:Ncollapse} shows that the continuous and discrete solutions are in agreement. Moreover, it is evident from Eq. (\ref{eq:DtaubNsq}) that the position of the charge for which the translocation time is minimized depends solely on the value of $\gamma$. The values of the fractional monomer position $\left(M/N\right)_\mathrm{min}$ measured from the leading chain end for which the translocation time is minimum, obtained by maximizing the second term on the right hand side of Eq. (\ref{eq:DtaubNsq}), are $0.34$, $0.40$ and $0.50$ for ideal, self-avoiding and rodlike chains, respectively. Clearly, the value of $\left(M/N\right)_\mathrm{min}$ is determined by entropic effects alone. The optimum location of the charge close to the middle of the chain can be rationalized as follows. Once the charged segment has been transported across the membrane to the \emph{trans} side, the likelihood of a reverse transition bringing the charge back to the \emph{cis} side is greatly reduced, owing to the free energy penalty associated with the reverse transition. Therefore, following the transport of the charge across the pore, the translocation of the remainder of the chain reduces to that of a shorter chain of equivalent length migrating across the membrane. The presence of the charge effectively divides the translocation process into two stages, namely, the translocation of $M$ polymer segments prior to the migration of the charge, and that of a polymer of length $N-M$ following the migration of the charge. The translocation time is thus minimized when the two stages of translocation before and after the transport of the charge through the pore require roughly the same time. Prior to the migration of the charged segment, there is a larger entropic barrier to translocation . As a result, translocation may be expected to be slower during the first stage involving $M$ segments than in the second stage involving the remaining $N-M$ segments. We hypothesize that the optimum position of the charge is such that the net entropic cost of translocation is minimized.

\begin{figure}
\centering
\resizebox{83mm}{!}{\includegraphics{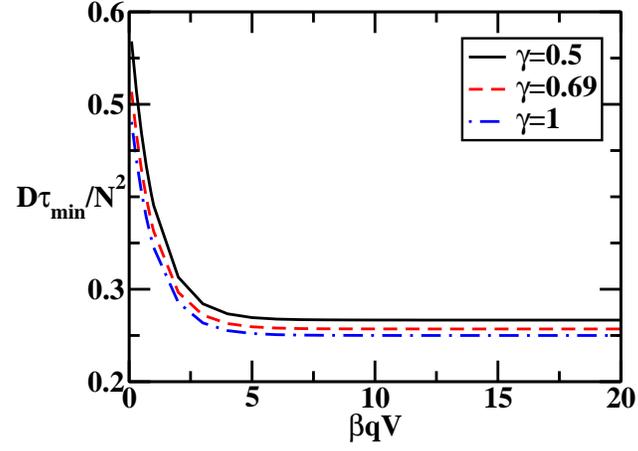} } 
\caption{Minimum value of the rescaled translocation time $D\tau/N^2$ as a function of $\beta qV$ with $\beta\epsilon=0$ and $\theta=0.5$ for an ideal chain (solid line), a self-avoiding chain (dashed line) and a rodlike chain (dashed-dotted line), each of $N=100$ segments. \label{fig:bqv}}
\end{figure}

The entropic cost of translocation of an ideal chain exceeds that of a self-avoiding walk, which in turn is associated with a greater entropic penalty than a rodlike chain [cf. Eqs. (\ref{eq:betaF})--(\ref{eq:betaFkN})]. Correspondingly, we expect the minimum translocation time to be the greatest for an ideal random walk, and the smallest for a rodlike chain. The position of the minimum is unaffected upon varying $\beta qV$, although the translocation time decreases when $\beta qV$ is increased, as expected. Figure \ref{fig:bqv} illustrates the asymptotic decrease of the minimum translocation time obtained from the discrete model with increase in $\beta qV$ for ideal, self-avoiding and rodlike chains, each of $N=100$ segments, with $\theta=0.5$.

\begin{figure}
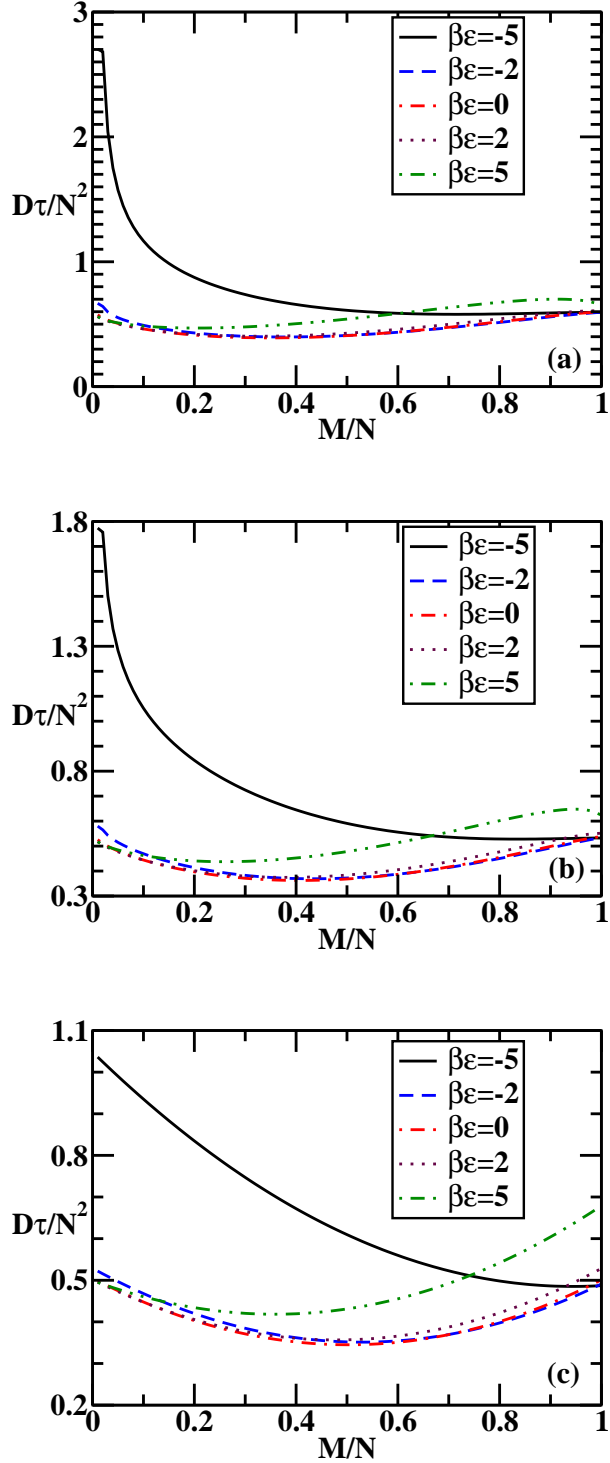

\centering
\resizebox{80mm}{!}{\includegraphics{irweps3.eps} } 
\vskip 2.25em
\centering
\resizebox{80mm}{!}{\includegraphics{saweps3.eps} } 
\vskip 2.25em
\centering
\resizebox{80mm}{!}{\includegraphics{rodeps3.eps} }
\caption{Rescaled translocation time $D\tau/N^2$ as a function of the fractional charge position $M/N$ for several values of $\beta \epsilon$ at $\beta qV=1$ and $\theta=0.5$ for (a) an ideal chain, (b) a self-avoiding chain, and (c) a rodlike chain, each possessing $N=100$ segments. \label{fig:eps} }
\end{figure}

\begin{figure}
\centering
\resizebox{80mm}{!}{\includegraphics{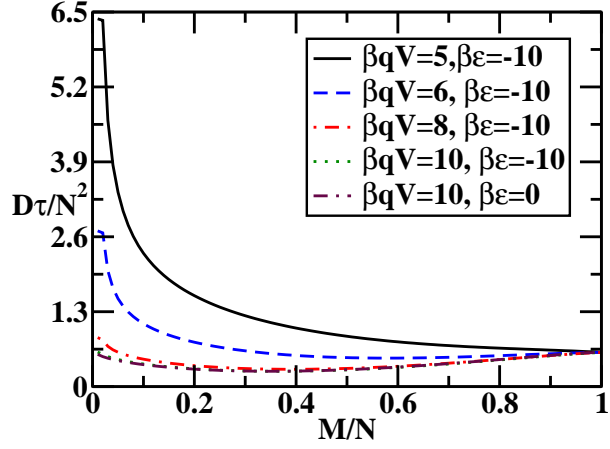} } 
\caption{Rescaled translocation time $D\tau/N^2$ as a function of the fractional charge position $M/N$ for several values of $\beta qV$ and with $\beta \epsilon=-10$ and $\theta=0.5$ for an ideal chain of $N=100$ segments. Also shown for comparison is the corresponding result at $\beta qV=10$ and $\beta\epsilon=0$. \label{fig:largeqeps} }
\end{figure}

\begin{figure}
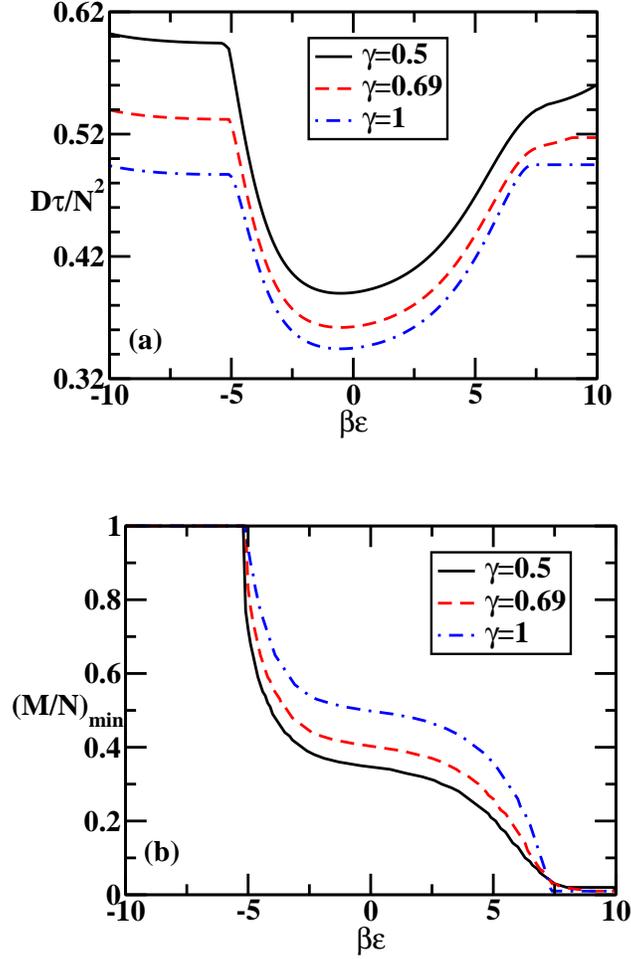

\centering
\resizebox{78mm}{!}{\includegraphics{tbeps.eps} } 
\vskip 2.5em
\centering
\resizebox{83mm}{!}{\includegraphics{Mbeps.eps} }
\caption{(a) Minimum value of the rescaled translocation time $D\tau_\mathrm{min}/N^2$ and (b) the fractional charge position $\left(M/N\right)_\mathrm{min}$ at which the translocation time is minimum as a function of $\beta \epsilon$ at $\beta qV=1$ and $\theta=0.5$ for an ideal chain (solid line), a self-avoiding chain (dashed line) and a rodlike chain (dashed-dotted line) of $N=100$ segments each. \label{fig:tmineps} }
\end{figure}

We next consider the effect of pore--charge interactions by introducing a non-zero $\epsilon$ in Eqs. (\ref{eq:betaF})--(\ref{eq:betaFkN}). The results from the discrete model are illustrated in Fig. \ref{fig:eps} for several values of $\beta\epsilon$ and with $\beta qV=1$ for ideal, self-avoiding and rodlike chains, and in Fig. \ref{fig:largeqeps} for an ideal chain at several values of $\beta qV$ with $\beta\epsilon=-10$, corresponding to strongly attractive interactions. We note that although we continue to employ the scaled translocation time $D\tau/N^2$, the collapse of the scaled translocation time for different chain lengths occurs only in the absence of interactions. It is clear that strong interactions, whether attractive or repulsive, slow down the translocation process. The effect of increasing the magnitude of $\beta\epsilon$ at fixed $\beta qV$ is similar to that of decreasing $\beta qV$ with $\beta\epsilon$ held fixed. Strongly attractive interactions that greatly exceed the voltage difference have a dramatic effect when the charge is located close to the leading chain end, and significantly slow down the translocation process. The translocation time is little affected by interactions when $\beta\epsilon$ is of the same order of magnitude as $\beta qV$. These features are manifest in Fig. \ref{fig:tmineps}, which also suggests a slight decrease in the minimum translocation time in the presence of weak pore--charge attraction [$ \left|\beta\epsilon \right| \lesssim O(1)$] relative to translocation through an inert pore. 

Figure \ref{fig:tmineps} demonstrates that the optimum charge position in the presence of strong pore--charge attraction is shifted towards the trailing end of the chain, whereas strong pore--charge repulsion effects a shift in the optimum charge position towards the leading chain end. Strongly attractive interactions hasten the reverse transition rate of the charged site, and are least effective in impeding translocation when the charge is situated at the trailing chain end. When the charge is situated near the leading end, pore--charge attraction greatly slows down translocation because of the increased tendency of the charge towards reverse migration. This effect is minimized when the charge is located at the trailing end. On the other hand, repulsive interactions slow down the reverse transition rate of the charged monomer and, hence, serve to impede the backward motion of the chain once the charge has been transported across the membrane. However, since a reflecting boundary condition is imposed at $k=N$, no reverse transition is allowed after the last segment has migrated, regardless of whether or not the final segment is charged. Furthermore, pore--charge repulsion causes an increase in the net free energy barrier to translocation, rendering charged locations in the middle of the chain unfavorable. As a result, pore--charge repulsion renders the leading chain end the most favorable location for the charge, although the leading end is only slightly more favorable than the trailing end (cf. Fig. \ref{fig:eps}). 

\section{\label{sec:twocharge} Extension to Multiple Charged Sites}

In this section, we consider the translocation of a polymer bearing $n$ charges of magnitudes $q_1,\ q_2,...,\ q_n$ at positions $M_1,\ M_2,...,\ M_n$ from the leading end through an inert pore. Analogous to Eqs. (\ref{eq:betaF})--(\ref{eq:betaFkN}), the instantaneous free energy of the translocating chain is
\begin{equation}
\beta F_{k} =
\begin{cases}
\left( 1-\gamma \right) \ln \left[ k\left( N-k\right) \right] 
\text{ for }1\leq k<M_{1}, \\
\left( 1-\gamma \right) \ln \left[ k\left( N-k\right) \right] -\beta
q_{1}V\text{ for }M_{1}\leq k<M_{2}, \\
....... \\
\left( 1-\gamma \right) \ln \left[k\left( N-k\right) \right] -\beta
\left( q_{1}+q_{2} +...+ q_n\right) V\text{ for }M_{n}\leq k\leq N-1 ,
\end{cases} \label{eq:betaFgen}
\end{equation}
with 
\begin{equation}
\beta F_{0}=\left( 1-\gamma \right) \ln N
\end{equation}%
and 
\begin{equation}
\beta F_{N}=\left( 1-\gamma \right) \ln N-\beta \left( q_{1}+q_{2}+...+q_n\right) V. \label{eq:betaFkNgen}
\end{equation}

The general solution for arbitrary $n$ may be obtained from the substitution of Eqs. (\ref{eq:betaFgen})--(\ref{eq:betaFkNgen}) into Eqs. (\ref{eq:uk})--(\ref{eq:fptdiscrete}). As an illustration, we explicitly consider only the continuous limit of translocation through an inert pore for the case $n=2$, whereby Eqs. (\ref{eq:taucont}) and (\ref{eq:Phi}), in conjunction with Eqs. (\ref{eq:uk}), (\ref{eq:wk}) and (\ref{eq:betaFgen})--(\ref{eq:betaFkNgen}), yield the result
\begin{multline}
\frac{D\tau }{N^{2}}=\int_0^1 dy\frac{B(y,\gamma ,\gamma )}{\partial
B(y,\gamma ,\gamma )/\partial y} \\ 
-\left( 1-e^{-\beta q_1 V}\right) \left[ \left\{ B\left( \frac{M_{2}}{N}%
,2-\gamma ,2-\gamma \right) -B\left( \frac{M_{1}}{N},2-\gamma ,2-\gamma
\right) \right\}  \right. \\
\left. + e^{-\beta q_2 V} \left\{ B\left( 2-\gamma ,2-\gamma \right) -B\left( 
\frac{M_{2}}{N},2-\gamma ,2-\gamma \right) \right\} \right] B\left( \frac{M_{1}}{N}%
,\gamma ,\gamma \right)  \\
- \left( 1-e^{-\beta q_2 V} \right)  \left\{ B\left( 2-\gamma ,2-\gamma \right) -B\left( \frac{M_{2}}{N}%
,2-\gamma ,2-\gamma \right) \right\} B\left( \frac{M_{2}}{N},\gamma ,\gamma
\right). \label{eq:tau2charge}
\end{multline}

\begin{figure}
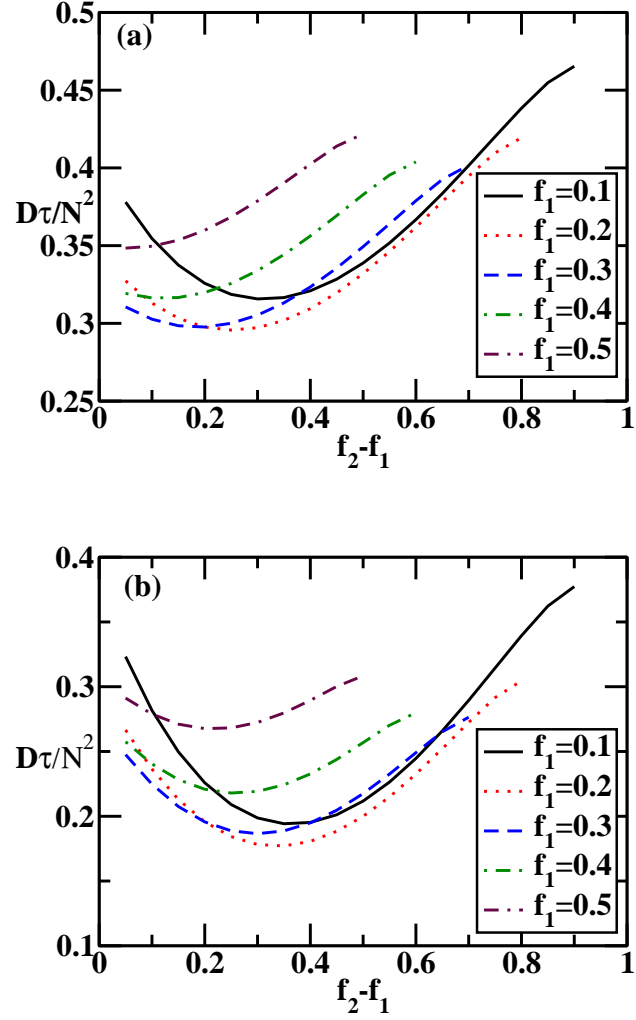

\centering
\resizebox{83mm}{!}{\includegraphics{tf2mf1q12ch.eps} } 
\vskip 2.5em
\centering
\resizebox{83mm}{!}{\includegraphics{tf2mf1q102ch.eps} }
\caption{Rescaled translocation time $D\tau/N^2$ as a function of the distance between the two charges expressed as a fraction of the chain length, $f_2-f_1$, for several values of $f_1$ at (a) $\beta qV=1$ and (b) $\beta qV=10$ for an ideal chain of $N=100$ segments. \label{fig:twocharget}}
\end{figure}

\begin{figure}
\centering
\resizebox{83mm}{!}{\includegraphics{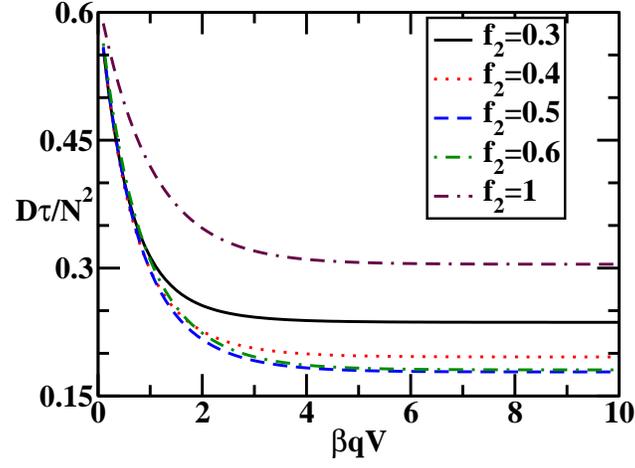} } 
\caption{Rescaled translocation time $D\tau/N^2$ as a function of $\beta qV$ with $f_1=0.2$ and at several values of $f_2$ for an ideal chain of $N=100$ segments. \label{fig:twochargetbqv} }
\end{figure}

Our findings are qualitatively similar to the results for the translocation of a chain with a single charge. Equation (\ref{eq:tau2charge}) confirms the scaling $D\tau \sim N^2$ for given $\gamma$, $\beta qV$, $f_1$ and $f_2$, where we have introduced the notation $f_1=M_1/N$ and $f_2=M_2/N$, under the assumption $q_1=q_2\equiv q$. Figures \ref{fig:twocharget} and \ref{fig:twochargetbqv} illustrate the results for the translocation time of an ideal chain of $N=100$ segments carrying charges at the fractional positions $f_1$ and $f_2$. The optimum charge positions now depend on the value of $\beta qV$, showing an increase in $f_2-f_1$ with increase in $\beta qV$. For an ideal chain, the optimum charge positions occur at $f_1\simeq 0.2$--$0.3$ and $f_2-f_1 \simeq 0.25$--$0.35$. Similar results (not shown here) are obtained for self-avoiding and rodlike chains, although the optimum charge locations are shifted to $f_1\simeq 0.3$--$0.35$ and $f_2-f_1 \simeq 0.25$--$0.3$, and $f_1\simeq 0.35$--$0.4$ and $f_2-f_1 \simeq 0.25$--$0.35$, respectively. Again, the minimum translocation time for given $\beta qV$ is longest for an ideal chain and shortest for a rodlike chain, and the translocation time asymptotically decreases with increase in $\beta qV$ for given $f_1$ and $f_2$.

A straight-forward extension of Eq. (\ref{eq:tau2charge}) for a polymer carrying $n$ charges yields the following expression for the translocation time:
\begin{multline}
D\tau =\int_{0}^{N}dk^{\prime }\left[ k^{\prime }\left( N-k^{\prime }\right) %
\right] ^{1-\gamma }\int_{0}^{k^{\prime }}dk^{\prime \prime }\left[
k^{\prime \prime }\left( N-k^{\prime \prime }\right) \right] ^{-\left(
1-\gamma \right) }- \\
\sum_{i=1}^{n}\left( 1-e^{-\beta q_{i}V}\right) \int_{0}^{M_{i}}dk^{\prime
\prime }\left[ k^{\prime \prime }\left( N-k^{\prime \prime }\right) \right]
^{-\left( 1-\gamma \right) } \\
\left[ \sum_{k=i}^{n-1}\left( \prod\limits_{j=i+1}^{k}e^{-\beta
q_{j}V}\right) \int_{M_{k}}^{M_{k+1}}dk^{\prime }\left[ k^{\prime }\left(
N-k^{\prime }\right) \right] ^{1-\gamma }+\left(
\prod\limits_{j=i+1}^{n}e^{-\beta q_{j}V}\right) \int_{M_{n}}^{N}dk^{\prime
}\left[ k^{\prime }\left( N-k^{\prime }\right) \right] ^{1-\gamma }\right].
\end{multline}

\section{\label{sec:sum} Summary and Conclusions}

In the present contribution, we investigate the voltage-driven translocation of inhomogeneously charged polymers through a nanopore. As a simple illustration of the effect of charge inhomogeneity, we first consider the translocation of a polymer bearing a single charge in the absence of pore--charge interactions. Our results reveal that the position of the charge minimizing the translocation time is determined solely by the nature of the polymer or, equivalently, the solvent quality, and is not necessarily situated in the middle of the chain as may be presupposed based on symmetry considerations alone. In fact, the symmetry is broken during protein translocation because a specific end of the protein is preferentially delivered to the membrane by the targeting machinery. Consequently, the two ends of the protein may be distinguished, with the leading end containing the signal sequence being adjacent to the pore at the start of the translocation process.

The minimum translocation time of an ideal chain is found to be the greatest, in consequence of the large entropic cost of translocation, whereas the minimum translocation time is least for a rodlike chain, which suffers no entropic penalty during translocation. The optimum charge position is found to lie at the midpoint of the chain for rodlike chains, and is shifted closer to the leading end for ideal and self-avoiding chains. This is because the presence of the charge effectively divides the translocation process into two stages, namely, the translocation of the portion of the chain preceding the charge and that of the chain segments following the charge. The portion of the chain preceding the charge must surmount a larger entropic barrier than that following the charge and, hence, moves relatively slower. The translocation time is expected to be minimum when the two stages of translocation require the same duration of time. The charge position leading to the minimum translocation time is therefore governed by the entropic cost of translocation.

In the presence of strong pore--charge interactions, both attractive and repulsive, the minimum translocation time increases relative to its value in the absence of interactions. The existence of strong attractive interactions shifts the optimum charge position towards the trailing chain end, whereas the optimum charge position lies at the leading chain end in the case of strong repulsive interactions. This observation is explained by the fact that whereas the reverse transition rate of the charged segment following its migration decreases in the case of repulsion, there is a large tendency for the charged segment to return to the pore immediately after it has been transported to the \emph{trans} side in the case of pore--charge attraction. 

We provide a simple illustration of the translocation of multiply-charged chains, and our results indicate that the effect of multiple charges on translocation through an inert pore is qualitatively similar to that of a single charge, and is also entropic in origin. Our results suggest possible mechanisms by which charge distribution may influence protein translocation across biological membranes. It is demonstrated that the translocation of proteins in biological systems may be accelerated by tuning pore--charge interactions and the distribution of charges on the chain. Moreover, our findings have potential technological implications for the design of peptides for medical applications such as drug delivery into cellular compartments.

\section*{Acknowledgments}

The authors would like to acknowledge support from the Welch Foundation (Grant No. C-1559) and the U.S. National Science Foundation (Grants No. CHE-0237105 and ECCS-0708765).

\bibliography{trans_paper}

\clearpage
\pagebreak

\clearpage
\pagebreak

\clearpage
\pagebreak

\clearpage
\pagebreak

\clearpage
\pagebreak

\clearpage
\pagebreak

\clearpage
\pagebreak

\clearpage
\pagebreak

\end{document}